\documentclass[pdflatex,sn-mathphys-num]{sn-jnl}

\usepackage{graphicx}%
\usepackage{multirow}%
\usepackage{amsmath,amssymb,amsfonts}%
\usepackage{amsthm}%
\usepackage{mathrsfs}%
\usepackage[title]{appendix}%
\usepackage{xcolor}%
\usepackage{textcomp}%
\usepackage{manyfoot}%
\usepackage{booktabs}%
\usepackage{algorithm}%
\usepackage{algorithmicx}%
\usepackage{algpseudocode}%
\usepackage{listings}%
\usepackage{tcolorbox}

\renewcommand{\l}{\mathopen{}\mathclose\bgroup\left}
\renewcommand{\r}{\aftergroup\egroup\right}

\let\originaleps=\epsilon
\let\epsilon=\varepsilon
\let\varepsilon=\originaleps
\usepackage{stmaryrd}

\theoremstyle{thmstyleone}%
\theoremstyle{thmstyletwo}%

\theoremstyle{thmstylethree}%

\begin{document}

\title[Article Title]{Detecting secondary-phase in bainite microstructure through deep-learning based single-shot approach}


\author{\fnm{Vinod} \sur{Kumar}}

\author{\fnm{Sharukh} \sur{ Hussain}}

\author{\fnm{Vishwas} \sur{ Subramanian}}

\author[*]{\fnm{P G} \sur{Kubendran Amos}}\email{prince@nitt.edu}
%

\affil[]{\orgdiv{Theoretical Metallurgical Group, Department of Metallurgical and Materials Engineering}, \orgname{National Institute of Technology Tiruchirappalli}, \orgaddress{ \postcode{620015}, \state{Tamil Nadu}, \country{India}}}

%


\abstract{Relating properties and processing conditions to multiphase microstructures begins with identifying the constituent phases.
In bainite, distinguishing the secondary phases is an arduous task, owing to their intricate morphology.
In this work, deep-learning techniques deployed as object-detection algorithms are extended to realise martensite-austenite (MA) islands in bainite microstructures, which noticeably affect their mechanical properties.
Having explored different techniques, an extensively trained regression-based algorithm is developed to identify the MA islands.
This approach effectively detects the secondary phases in a single-shot framework without altering the micrograph dimensions.
The identified technique enables scalable, automated detection of secondary phase in bainitic steels.
This extension of the detection algorithm is suitably prefaced by an analysis exposing the inadequacy of conventional classification approaches in relating the processing conditions and composition to the bainite microstructures with secondary phases. }




\maketitle

\section{Introduction}

Despite being hard to define and comprehend~\cite{aaronson1990bainite,aaronson2002progress}, bainite steels continue to be of huge interest to engineers~\cite{adamczyk2021studies,ramachandran2021role}. 
This interest is primarily driven by the wide range of properties they offer through their secondary phases~\cite{tomita1983improvement,edmonds1990structure}. 
The secondary phases, while yielding the desired set of properties, significantly increase the complexity of the bainite microstructure~\cite{zajac2005characterisation}.
Achieving the desired properties demands meticulous control over the type and volume fraction of the secondary phases in bainite steels~\cite{bleck2011improved}. 
A control over the nature of these secondary phases can only be gained through a proper understanding of the effect of composition and/or processing in bainite. 
Generally, the influence of composition and/or processing on a microstructure is largely evident. 
Consequently, a required phase-fraction is achieved by exposing an alloy of definite composition to a suitable processing treatment~\cite{kumar2018toughness}. 
However, owing to the intricacy of the bainite microstructure, understanding the effect of composition and processing treatments becomes challenging. 
Therefore, human experts are invariably involved to realise the features of the bainite microstructure, particularly the type and configuration of the secondary phases~\cite{zajac2005characterisation}. 
But recently, a paradigm shift is noticeable in the analysis of the microstructure with respect to its composition and processing conditions~\cite{durmaz2021deep,azimi2018advanced}. 
Lately, deep learning has increasingly been employed to realise the change induced in the intricate bainite microstructure by the processing history and/or chemical make-up~\cite{muller2020classification,muller2021microstructural,ackermann2022automated,kumar2025interpretable}.

A wide range of computer vision, machine learning, and deep learning approaches have been adopted to distinguish various steel microstructures, including bainite, pearlite, martensite, and ferrite. 
Realising the distinction among these steels is rather straightforward, owing to the characteristic differences in the morphology of the phases. 
Correspondingly, after adequate computer vision-based pre-processing, machine learning techniques like support vector machines are employed to classify the different microstructures of steels, even distinguishing between upper and lower bainite~\cite{muller2020classification,muller2021microstructural}.
However, when the intent is to relate the processing condition and/or composition to the microstructure, these classifications offer only a superficial understanding. 
In other words, the processing for a given steel rarely brings about a significant change in the type of the constituent phases; instead, it alters the phase-fraction and the morphology of the secondary phase(s). 
This means the adopted technique should be capable of distinguishing microstructures with the same combination of phases, primarily based on their intricate morphology and distribution. 
The inadequacy of conventional machine learning techniques in classifying such microstructures with similar phases necessitates further analysis, such as detecting the critical phase(s) that dictate the properties and are influenced by the chemical composition and/or processing conditions~\cite{kumar2024inadequacy}.

In bainite microstructures resulting from non-isothermal transformation, a secondary phase referred to as martensite-austenite (MA) islands is introduced, which establishes significant changes in the mechanical properties depending on their morphology and distribution. 
Therefore, any attempts to understand the effect of composition and/or processing conditions on bainite microstructures containing MA islands demand realising this characteristic secondary phase~\cite{xu2021significant,jiang2021tempering,lu2023structure}. 
This necessity arises from the assumption that conventional machine-learning-based classification would be insufficient. 
To that end, recently, a masked region-based convolutional neural network (RCNN) within the framework of Detectron2 has been employed to segment MA islands~\cite{ackermann2022automated}. 
This work, while pioneering the effort in realising the MA islands in intricate bainite microstructures, still offers scope for further refinement. Firstly, the approach involves a combination of classification and detection for realising the MA islands. 
Secondly, the accuracy of the detection is dependent on the size of the micrographs, with the models offering greater precision when the images are smaller than the original. 
The former component of the study indicates the computational demand, whereas the latter limits the general usability of the approach.
Given these apparent limitations, in this work, a single-shot tool is developed to efficiently detect the MA islands in the bainite microstructure. Before expose various techniques to identify an efficient approach for realising the secondary phase, the bainite microstructures are analysed through machine-learning tools to expose the inadequacy of the conventional treatment and to establish the need for detecting the critical MA islands.

\section{Overall Workflow}

\begin{figure}[h]
    \centering
    \includegraphics[width=\linewidth]{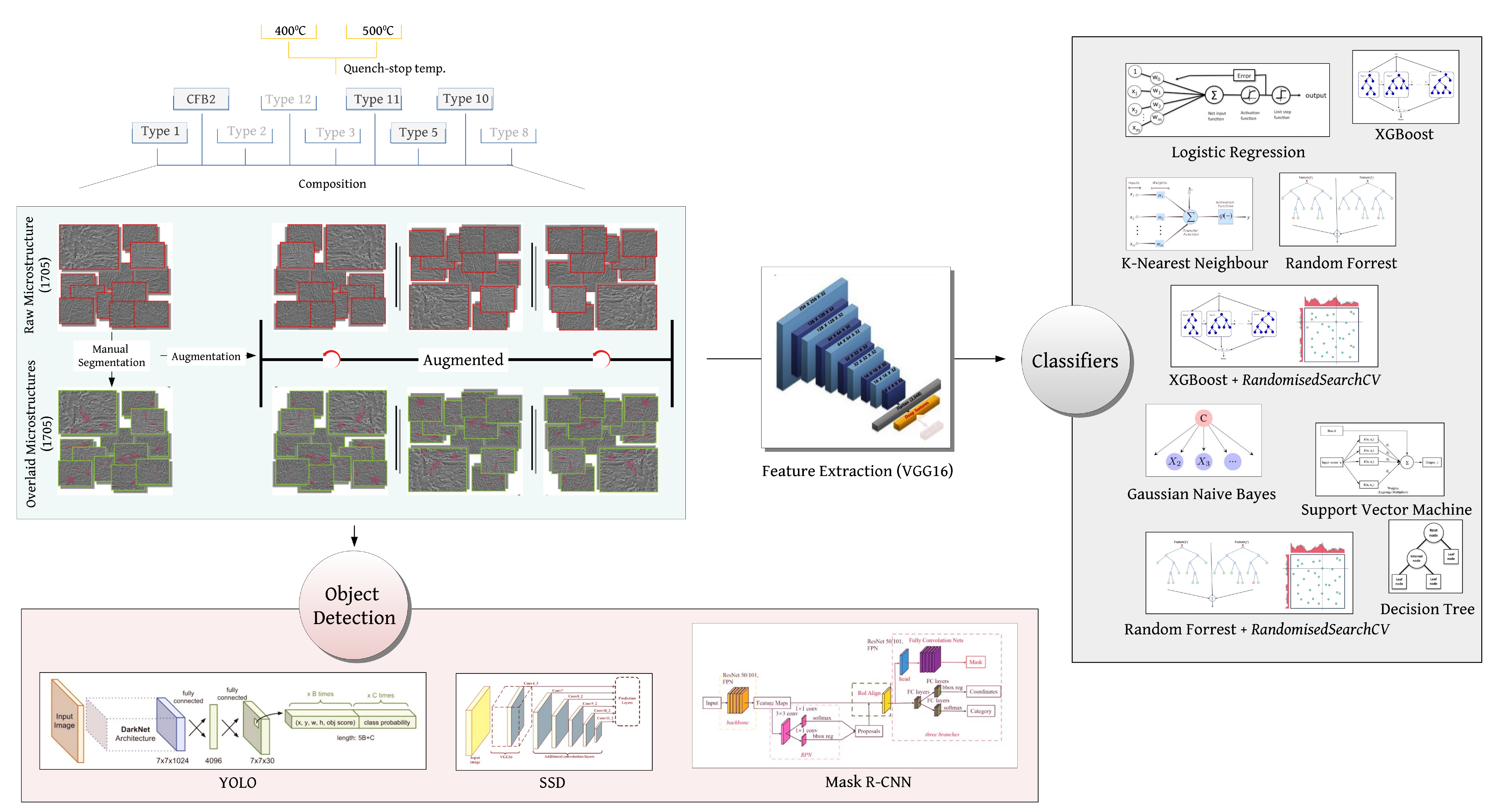} 
    \caption{The overall work flow of the present investigation including processing of the original bainite microstructure, feature extraction, classification using range of machine-learning techniques and development of secondary-phase detection models.  
    }
    \label{fig:flow}
\end{figure}

\subsection{Bainite Dataset}

The present model to detect MA islands in bainite microstructures is developed by training on the known dataset published in ref~\cite{iren2021aachen}. 
This dataset includes micrographs of nine different steels distinguished by their composition, and exposed to two different heat treatment cycles. Of the nine, only five steels are considered in this work (CFB2, Type-1, -5, -10, and -11), given the number of microstructures associated with each type. This consideration yields a set of 1476 bainite micrographs with 8539 MA islands in them. All steel types, before being hot rolled at 900~$^0$C, are austenised at 1200~$^0$C for 10 minutes. During hot rolling, a strain of 0.3 is applied at a rate of 10s$^{-1}$. The difference in the processing is introduced by cooling some samples, across compositions, at the rate of 5Ks$^{-1}$ and others at 0.3Ks$^{-1}$ to the quench-stop temperatures of 400$^0$C and 500~$^0$C, respectively. These disparities in the cooling rate and quench-stop temperature are treated as the distinguishing processing conditions in the present investigation. High quality images of the bainite microstructures from different steels are obtained through Scanning Electron Microscopy (SEM) at 4000$\times$ magnification. A duplicate of the 1476 bainite micrographs, called the overlaid dataset, is created by superimposing the polygons that highlight MA islands. The coordinates of the polygons are assigned by human experts after careful consideration. Therefore, besides the raw dataset of 1476 bainite micrographs, there exists a duplicate overlaid dataset highlighting the MA islands.

\subsection{Deep-learning approach}

The entirety of the current work is graphically presented in Fig.~\ref{fig:flow}. This workflow can be broadly categorised into two parts. In the preliminary analysis, the bainite microstructures containing MA islands are classified through a wide range of machine-learning models to explicate the inadequacy of classification in realising the effect of processing condition and composition. Instead of the micrographs themselves, the characteristic feature vectors, extracted through the VGG16 network, are classified using various models. The size of the dataset is essentially tripled, to enhance the performance of the classifiers, through augmentation involving flipping and rotation. Besides the raw images, overlaid microstructures with highlighted MA islands are also augmented and classified.

\subsection{Suite of classifiers}

The preliminary investigation indicating the need for detecting MA islands employs nine machine-learning algorithms to classify the VGG16-extracted deep feature vectors. Although the overview of the models along with the different hyperparameters used are discussed elsewhere (supplementary document), the reason for the choice of these models is concisely presented here.

Given its proven ability to distinguish various microstructures of steels, including upper and lower bainite~\cite{muller2020classification,muller2021microstructural}, Support Vector Machines (SVMs) is included in the suite of classifiers.
For distinguishing intricate morphologies SVMs with a radial basis function (RBF) kernel are employed. 
The regularization parameter \texttt{C} was set to 1.0 to balance margin maximization and misclassification.

Logistic regression serves as a baseline model, providing a reference for evaluating the non-linear separability in the feature vectors, while the Gaussian Naive Bayes algorithm, a probabilistic model, is considered to realise any distribution of the feature vectors aiding in classification.
The regularization strength is set at \texttt{C = 1.0}, utilizing the \texttt{lbfgs} solver with L2 regularization to manage complexity, and for Gaussian Naive Bayes algorithm, variance smoothing is considered to be \texttt{1e-9}.

Focusing on the potential clustering of the feature vectors in the multidimensional space, the k-nearest neighbour (kNN) algorithm classifies the microstructures through a non-parametric, instance-based learning, offering a robust check against overfitting. In this treatment, five neighbors (\texttt{n\_neighbors = 5}) were considered with uniform distance weighting.

Given the heterogeneity and non-linearity in the deep vectors introduced by the intricate morphology of the microstructures, decision-tree-based classifiers are included, with their generality enhanced by random forest through aggregation of predictions from an ensemble of decorrelated trees. The performance of random forest is separately enhanced by hyperparameter optimisation using RandomizedSearchCV. As a strong alternate to the tree-based classifier, Extreme Gradient Boosting (XGBoost) is included, where the morphological differences in the MA islands are realised through a gradient-based approach. Hyperparameters of the XGBoost are also efficiently tuned by RandomizedSearchCV and considered as a distinct algorithm. In decision-tree classifier, the maximum depth is left unrestricted (\texttt{max\_depth = None}), and a minimum of two samples (\texttt{min\_samples\_split = 2}) is  split, using the \texttt{Gini} impurity criterion to guide node division. Each forest comprises of 100 trees (\texttt{n\_estimators = 100}), with unrestricted tree depth (\texttt{max\_depth = None}) and a \texttt{sqrt} setting for \texttt{max\_features} to ensure a randomized selection of features at each split. 
Random Forest models are additionally optimized using \texttt{RandomizedSearchCV} over a five-fold cross-validation including a hyperparameter search \texttt{n\_estimators} between 100 and 200, \texttt{max\_depth} ranging from 4 to 16, \texttt{min\_samples\_split} between 2 and 10, and \texttt{max\_features} set to either \texttt{sqrt} or \texttt{log2}. 
XGBoost operates with parameters \texttt{n\_estimators = 100}, \texttt{learning\_rate = 0.3}, and \texttt{max\_depth = 3}. 
The search space for its hyperparameters encompasses texttt{n\_estimators} ranging from 60 to 200, \texttt{learning\_rate} between 0.01 and 0.3, and \texttt{max\_depth} from 3 to 10. Moreover, parameters such as \texttt{subsample} (0.6 to 1.0), \texttt{colsample\_bytree} (0.6 to 1.0), and \texttt{gamma} (0 to 5) is varied.

The nine different models constituting the present algorithm suite are trained to classify the bainite microstructures with MA islands based on composition and quench-stop temperature, \textit{id est} processing condition, separately. The performance of these classifiers is monitored to assess the need for separately realising the MA islands. Put simply, a convincing classification of the feature vectors by the algorithms establishes a relation between the microstructures and the composition or processing condition, which can be examined to understand the effects of these factors on the bainite micrograph, including the morphology of the secondary phases. On the other hand, an inaccurate classification demands deeper investigation of the bainite microstructural features, particularly the MA islands.

\subsection{Object-detection algorithms}

Given the intricacy of the bainite microstructure, and the influence of the MA islands on the mechanical properties, the current study primarily focuses on realising a deep-learning technique for single-shot detection of this secondary phase without compromising the original dimensions of the micrographs. The automated segmentation of the MA islands, in the existing approach, indicates an increase in accuracy by sectioning the micrograph and integrating classification to enhance efficiency. The present treatment, on the other hand, relies on the robust training of detection models through manual labelling. Stated otherwise, treating the overlaid data as the ground truth, all the 8539 MA islands in the 1476 raw microstructures are manually labelled to develop a single-shot detection model. The original annotations are transformed in-keeping with the applied augmentation by updating the bounding box coordinates to maintain spatial consistency.

Detection of the MA islands, in the present work, is investigated in three different frameworks including YOLO, Single Shot Multibox Detector (SSD), and Mask R-CNN. Through a regression-based prediction of the bounding boxes, YOLO offers a potentially high-speed detection of the MA islands in the microstructures. SSD, while ascertaining the bounding boxes that encompass secondary phases using feature maps, is capable of detecting objects of varied sizes, including small ones. Despite being a two-stage detector, Mask R-CNN, which has been used in the existing work, extends Faster R-CNN by adding a parallel branch for pixel-level instance segmentation. A detailed description of these models along with their respective hyperparameters is discussed elsewhere (supplementary document). Even though all these algorithms are trained and tested on original micrographs without altering the dimension, their underlying architecture lends itself to scalable microstructural analysis.

The YOLO model is initialized with pre-trained COCO weights, fine-tuned with a learning rate of \texttt{0.001}, momentum \texttt{0.937}, and weight decay \texttt{0.0005}. Anchor boxes are automatically optimized during training to adapt to the size distribution of MA islands. A batch size of 16 and input image size of 640\,\texttt{px} were used without altering the original SEM resolution.
A learning rate of \texttt{0.001}, weight decay \texttt{0.0005}, and a batch size of 32 are applied for SSD. Default boxes (anchors) are adapted to better match the typical scale of MA islands. 
In Mask R-CNN, ResNet-50 with Feature Pyramid Networks (FPN) backbone is used, employing a learning rate of \texttt{0.0025}, weight decay \texttt{0.0001}, and batch size of 8. 

\section{Result and Discussion}

\subsection{Classification of bainite microstructure}

\begin{figure}[h]
    \centering
    \includegraphics[width=\linewidth]{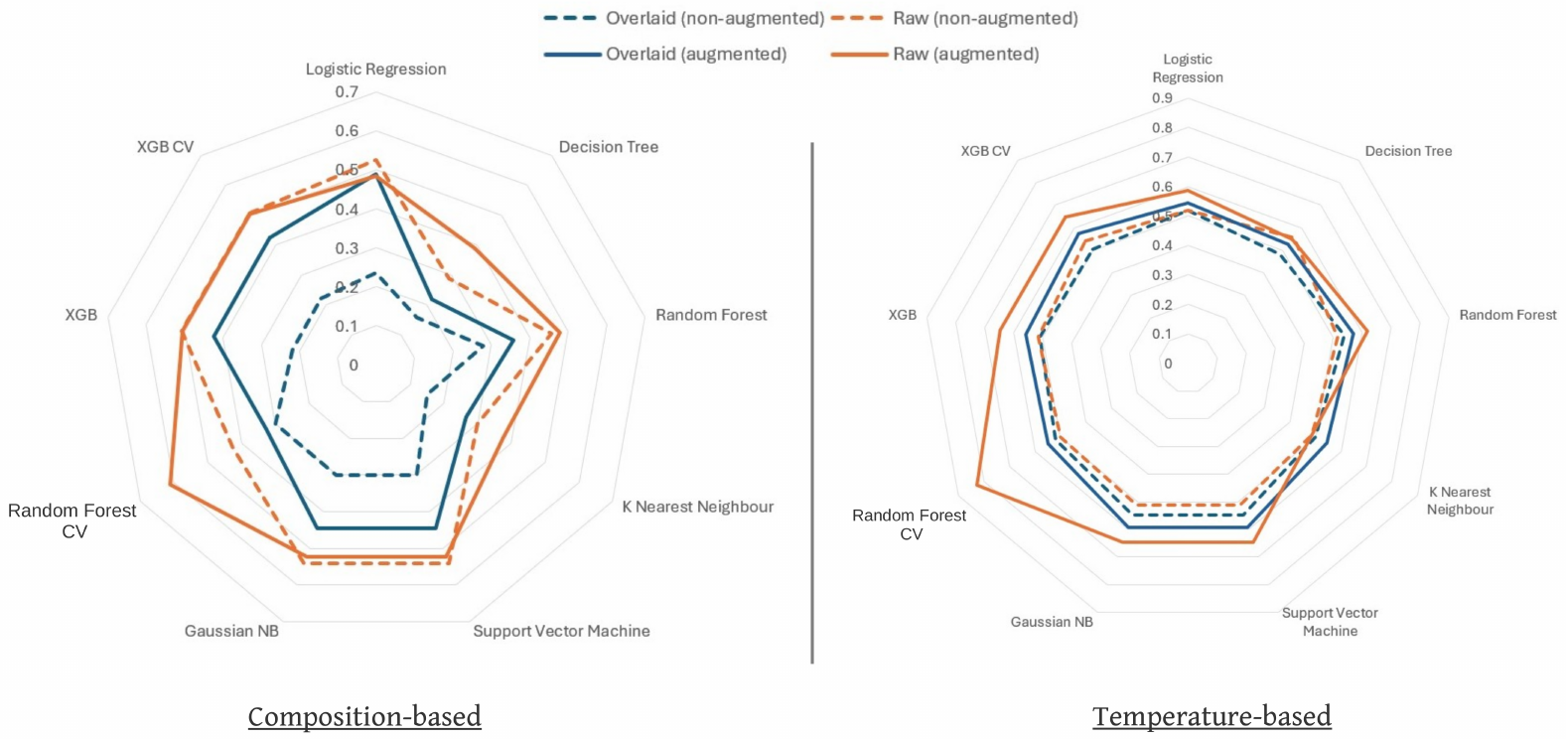} 
    \caption{Performance of various classifier on the original and augmented dataset of raw and overlaid bainite microstructures. 
    }
    \label{fig:class2}
\end{figure}

The overall accuracy of the different algorithms in classifying bainite microstructures based on their composition and quench-stop temperature is graphically represented in the left and right panels of Fig.~\ref{fig:class2}, respectively. The performance of the classifiers on both raw and overlaid microstructures, in their original and augmented forms, is included in this illustration.

In Fig.~\ref{fig:class2}, it is evident that the performance of all the models is noticeably better in temperature-based classification when compared to composition-based classification. This increased accuracy can be attributed to the significantly lesser number of classes for the given set of data. In other words, while 1476 microstructures are partitioned between two different quench-stop temperatures, they encompass five different compositions.

\subsubsection{Temperature-based}

\begin{figure}
    \centering
    \includegraphics[width=\linewidth]{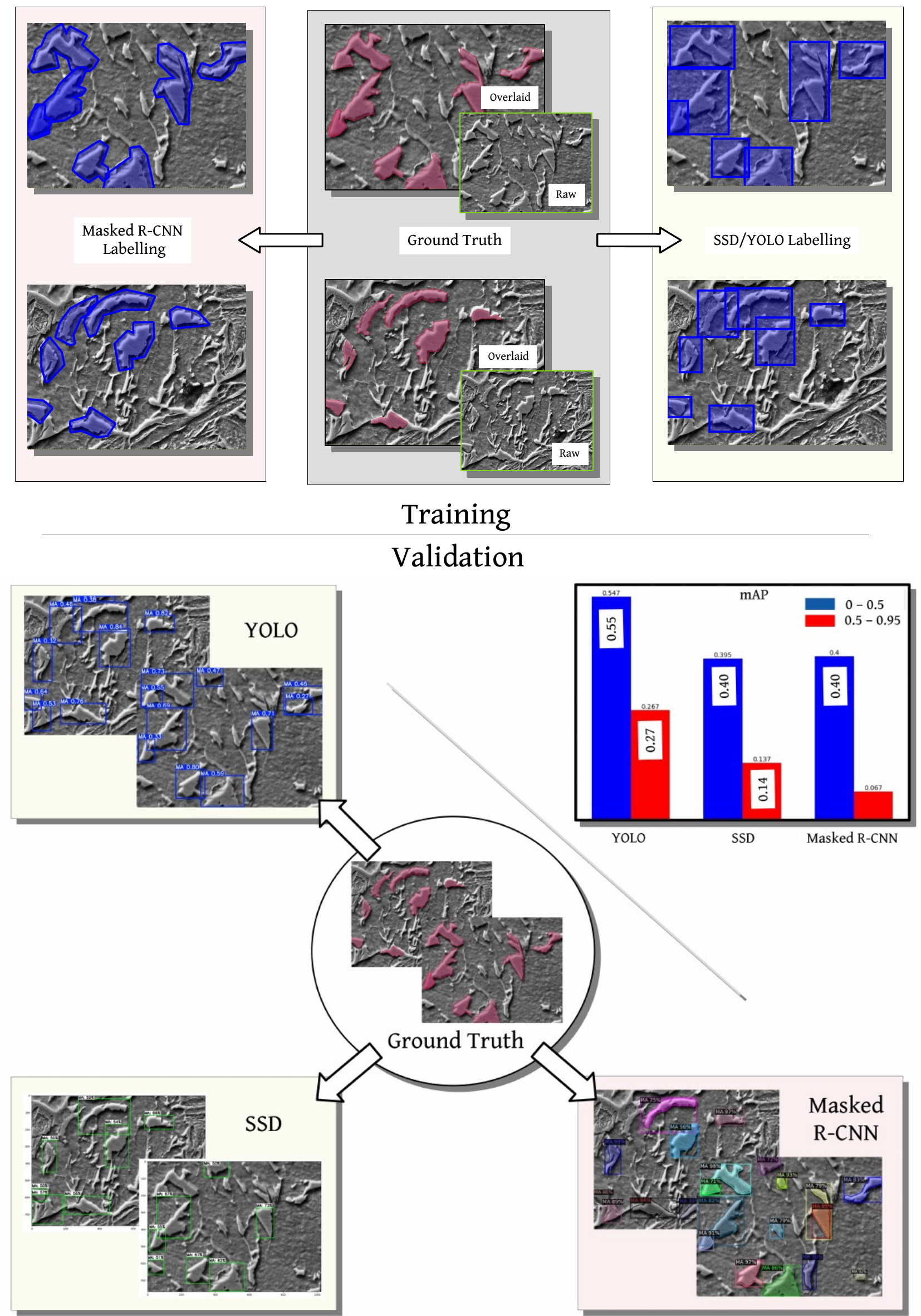} 
    \caption{Training and validation of the detection techniques on the raw microstructure. The performance of the different tools is included as mean Average Precision.  
    }
    \label{fig:trainVal}
\end{figure}

In the temperature-based classification, the difference in the accuracy between the original raw and overlaid micrographs, though existent, is not significant. This marginal difference in the classification accuracy of raw and overlaid unaugmented microstructures indicates that highlighting MA islands does not explicate the distinction in the overall bainite micrographs introduced by the varied quench-stop temperature. The accuracy of the temperature-based classification of the overlaid micrographs is not noticeably enhanced with augmentation. This marginal increase in performance indicates that the intrinsic distinguishing features between the two quench-stop temperatures are already well-represented in the original dataset, and the augmentation does not contribute any significantly new discriminative information. All models perform similarly across the augmented raw microstructures, except for random forest with RandomizedSearchCV, which exhibits an accuracy of 0.83 in classifying based on quench-stop temperature.

\subsubsection{Composition-based}

Unlike temperature-based analysis, the performance of each model varies noticeably when the classification is focused on composition. This trend is noticeable in the augmented and original datasets of both raw and overlaid bainite microstructures. The subtle and overlapping differences in the feature vectors, in the composition-based classification, influence the accuracy of the models. This influence varies with the classifier, depending on its ability to realise non-linear and complex interactions between the features. 

Irrespective of the varied performance, it is evident in Fig.~\ref{fig:class2} that the accuracy of composition-based classification significantly improves with augmentation, as opposed to temperature-based classification. This increase in accuracy with augmentation indicates the under-representation of the effect of composition in the bainite microstructure. Augmentation, in essence, brings out the morphological changes introduced by composition, thereby facilitating enhanced classification. Similar to the temperature-based investigation, random forest with RandomizedSearchCV offers the maximum accuracy in classifying the bainite microstructures based on composition. The stand-out performance of random forest CV, in both composition- and temperature-based classification, suggests that both these parameters introduce an intricate and largely overlapping change to the bainite microstructure. 

Despite the relatively better performance, the accuracy of random forest CV reduces from 0.83 to 0.61 in the composition-based classification. Furthermore, it is worth noting that the performance of classifiers is, in general, better over raw microstructures when compared to overlaid microstructures. This trend indicates that the accuracy of classification is enhanced when equal importance is extended to all features, as opposed to an additional emphasis on the MA islands characterising the overlaid microstructures. Stated otherwise, apparently all the features of the bainite microstructures, along with the MA islands, are influenced by the quench-stop temperature and composition, and a holistic consideration yields better performance in classification.

\subsection{Detection of MA islands}

The preliminary investigation, though unravelling a few critical aspects of the bainite microstructures with MA islands, is not completely adequate to isolate the effect of quench-stop temperature, \textit{id est} processing condition, and composition. Therefore, considering its noticeable influence, the analysis is extended to developing one-shot techniques for identifying the secondary-phase MA islands in the micrographs.

The training of the detection algorithms is illustrated in the top window of Fig.~\ref{fig:trainVal}. Of the overall dataset, both original and augmented, 80$\%$ is set aside for training and the rest for testing/validation. The detection models, irrespective of their underlying architecture, are trained to detect MA islands through manual labelling. In YOLO and SSD, the MA islands in the bainite microstructures are annotated through bounding boxes, whereas polygons exactly encompassing the secondary phase are used in Mask R-CNN. Highlighted sections in the overlaid microstructures serve as the ground for manual labelling.

\subsubsection{Raw dataset (Original)}

The performance of the different detection techniques, after adequate training, on the unseen bainite microstructures with MA islands is shown in the lower window of Fig.~\ref{fig:trainVal}. Though each technique adopts its own metrics to describe performance, mean average precision (mAP) is employed to facilitate the understanding of relative accuracy.  The mAP of YOLO, SSD, and Mask R-CNN are collectively presented in Fig.~\ref{fig:trainVal}. Given the models detect only one class of object in the bainite microstructures, \textit{id est} MA islands, the mAP and average precision are identical. This metric is determined from the area enclosed by the precision-recall curve, where precision describes the (true) positive detections in view of inaccurate (or false) realisations, while recall considers the accuracy with respect to the ground truth. 

It is evident in Fig.~\ref{fig:trainVal} that YOLO outperformed both SSD and Mask R-CNN in detecting the MA islands in bainite microstructures. For 0.5 IoU (Intersection over Union), YOLO achieves a mAP score of 0.55, whereas 0.4 is the accuracy of both SSD and Mask R-CNN. The end-to-end grid-based detection, along with the inherent optimisation of the anchors, offers an edge to YOLO when compared to other detection techniques. Fig.~also indicates that the compromised performance of the models is primarily due to the increased false predictions, which compromise the overall mAP score.

\subsubsection{With Augmentation}

\begin{figure}
    \centering
    \includegraphics[width=\linewidth]{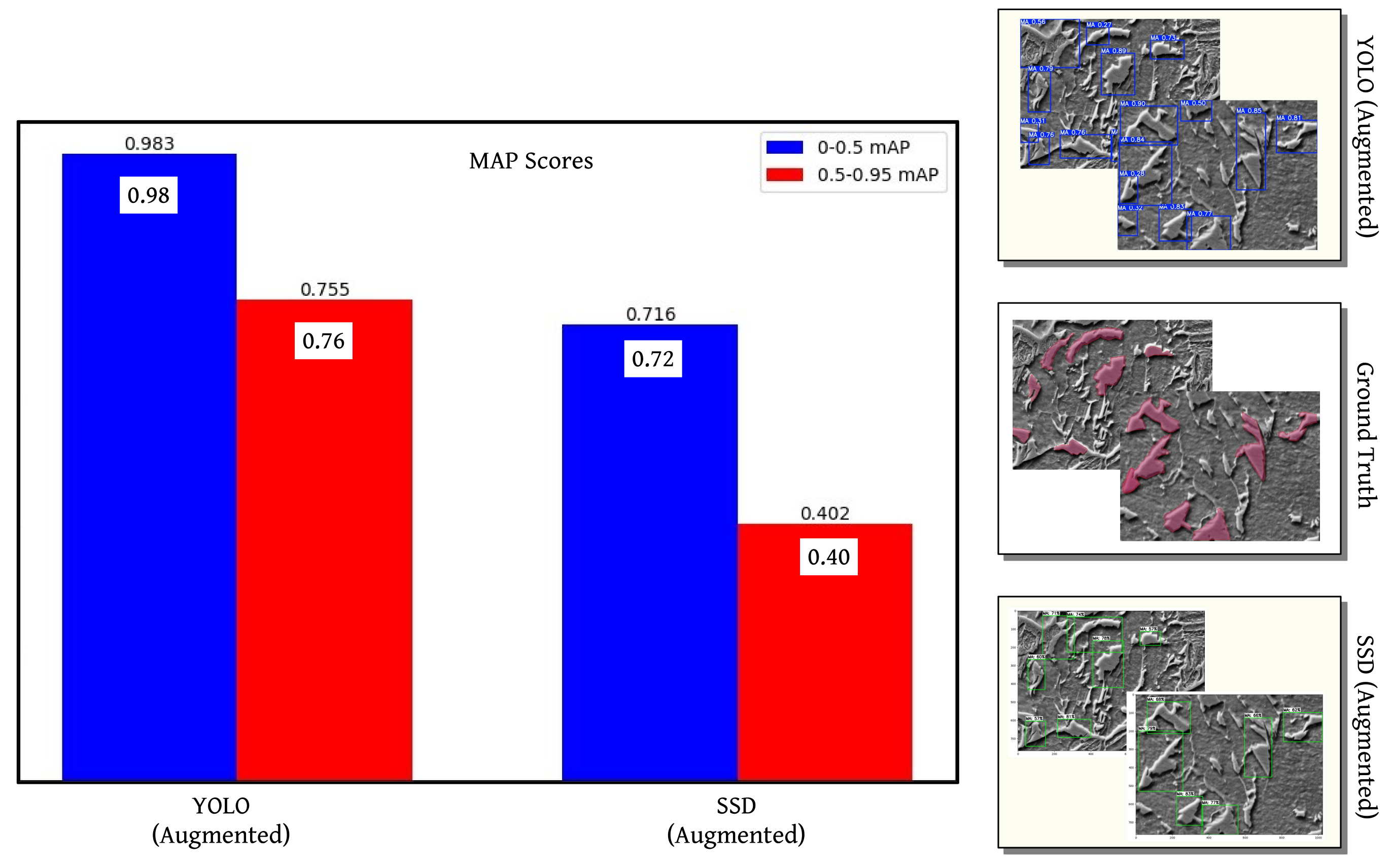} 
    \caption{a) The enhanced performance of YOLO and SSD owing to the training on the augement set of bainite microstructure.
    }
    \label{fig:Aug}
\end{figure}

As shown in Fig.~\ref{fig:Aug} , though each algorithm detects the MA islands with varying accuracy, the performance of none is sufficiently adequate to be treated as a single-shot detection tool. Therefore, two of the three detection models are trained on the augmented dataset of raw microstructures. By significantly increasing the dataset, YOLO and SSD are exposed to three-times more micrographs. Given the complexity in annotating the secondary phases and its performance over the original dataset, Mask R-CNN is not trained over the augmented set of micrographs.

The performance of YOLO and SSD, after training over the augmented dataset of bainite microstructures, is shown in Fig.~\ref{fig:Aug}. The performance of both techniques noticeably increases upon learning from the augmented dataset. While SSD achieves a mAP score of 0.72, with mAP (0.5 IoU) = 0.98, YOLO exhibits almost perfect detection of MA islands. In other words, the false detections by the YOLO algorithm significantly reduce with training on the augmented dataset, ultimately rendering an appropriate model for the single-shot detection of MA islands.

\section{Conclusion}
Realising critical secondary phases in the bainite system generally demands significant attention from experts.
If detection in itself is a demanding task, one can only imagine the efforts required to comprehend the changes introduced by composition or processing conditions in these intricate phases.
Similar effort is also required when attempts are made to relate the bainite microstructure to properties.
To address these concerns, in the present work, a regression-based object detection algorithm is extended?after considering other techniques?to realise a critical secondary phase known as martensite-austenite (MA) islands.
This approach facilitates accurate single-shot detection of the intricate MA islands without the need to alter the dimensions of the bainite microstructure.
The realisation of the present algorithm can be further extended to unravel the effects of composition and processing conditions on secondary phases.
In subsequent work, attempts will be made to extend this approach to other secondary phases beyond MA islands.


\end{document}